\documentstyle[12pt]{article}
\begin{document}
\begin{center}
{\LARGE {\bf Duality, Time-asymmetry and the Condensation of Vacuum}}\\ 
\vspace{1cm} 
${\bf H.Salehi^{1,2,3}~ and~ H.R.Sepangi^1}$\\ \vspace{0.5cm} 
{\small {$^{1}$ Department of Physics, Shahid Beheshti University, Evin,
Tehran 19834,  Iran.\\ $^{2}$ Institute for Studies in Theoretical Physics
and Mathematics, 19395-1795,  Tehran, Iran.\\ $^{3}$Institute for Humanities
and  Cultural Studies, 14155-6419, Tehran, Iran.}}
\end{center}
\begin{abstract}
A variant of the divergence theory for vacuum-condensation developed in a
previous communication  is analyzed from the viewpoint of a 'time'
asymmetric law  in vacuum. This law  is found to establish a substantial
distinction  between dynamically allowed vacuum-configurations related  by
signature changing duality transformations. \vspace{5mm}
\vspace{15mm}\\
\end{abstract}
\section{Introduction}
Vacuum-condensation is increasingly playing an important role in
Lorentz-invariant theories such as particle physics. The issue is, however,
somewhat controversial because there exists an apparent incompatibility between
vacuum-condensation and the requirement of the exact Lorentz-invariance.
Indeed, one of the defining characteristics of the exact Lorentz-invariance
in quantum field theory is a vanishing energy-density for vacuum. On the
other hand, any condensation of vacuum provides us, in general, with
characteristic mass-scales which can have a non-vanishing contribution to
the cosmological constant. This remark suggests that the origin of the
vacuum-condensation in Lorentz-invariant theories may conceivably be located
in external constraints reflecting a principal violation of
Lorentz-invariance. Lack of experimental evidences, however, makes it very
difficult to fully understand the physical content of Lorentz-noninvariant
theories. But, at the preliminary level one could analyze the violation of 
Lorentz-invariance in 
Einstein-Minkowski space in terms of intuitively clear
theoretical concepts which encode the expected characteristics of a
Lorentz-noninvariant vacuum. Here two theoretical concepts may be viewed as
a concrete proposal, namely the universal length $l_0$ and the related
internal vector $N_\mu $. The physical interpretation of the pair $%
(l_0,N_\mu )$ could be as follows\footnote{The idea 
of the universal length and  the related internal vector was
originally suggested in  \cite{blok} and subsequenly used in many 
contexts, e.g. in \cite{nielsen,salehi}}: The
universal length $l_0$ serves as a resolution limit for the physical
distances that can even be probed in realistic measurements. Clearly,
a resolution limit of this type is incompatible with the requirement of the
exact Lorentz invariance. In fact, it is only by relating $l_0$ to the
kinematical frame of a preferred inertial observer in vacuum that the
concept of the universal length makes sense. The internal vector $N_\mu $
serves as just characterizing the four-velocity of such a preferred inertial
observer and it must therefore be a time-like vector. In this way the pair 
$(l_0,N_\mu )$ arise as the immediate physical characteristics of a Lorentz
non-invariant vacuum.

\section{Divergence theory}

Given the pair $(l_0,N_\mu )$, we may analyze the structure of vacuum in
Einstein-Minkowski space in terms of the dynamical coupling of a real scalar
field $\phi $ to the internal vector $N_\mu $. Following \cite{salehi}, such a
dynamical coupling can be brought into a divergence form by defining the
current 
\begin{equation}
\label{eq1}J_\mu (\phi )=-\frac 12\phi \stackrel{\leftrightarrow }{\partial }%
_\mu \phi ^{-1}.
\end{equation}
It can be easily shown that 
\begin{equation}
\label{eq2}\partial _\mu J^\mu (\phi )=\phi ^{-1}(\Box \phi -\phi
^{-1}\partial _\mu \phi \partial ^\mu \phi ),
\end{equation}
and 
\begin{equation}
\label{eq3}J_\mu (\phi )J^\mu (\phi )=\phi ^{-2}\partial _\mu \phi \partial
^\mu \phi. 
\end{equation}
Combining equations (\ref{eq2}) and (\ref{eq3}), we obtain the identity 
\begin{equation}
\label{eq4}\Box \phi +\Gamma \{\phi \}\phi =0,\hspace{5mm}\mbox{with}%
\hspace{5mm}\Gamma \{\phi \}=-J_\mu (\phi )J^\mu (\phi )-\partial _\mu J^\mu
(\phi ).
\end{equation}

Henceforth, we refer to $\Gamma \{\phi \}$ as the dynamical mass term. Note
that 
equation (\ref{eq4}) is a consequence of  definition (\ref{eq1}) and not a
dynamical law for $\phi $. However, by specifying the source of the
divergence in the dynamical mass term, a large class of dynamical theories
can be constructed. We shall study a theory in which the source of the
divergence is specified by the relation 
\begin{equation}
\partial _\mu J^\mu (\phi )=l_0^{-1}N_\mu J^\mu (\phi ).
\label{eq5}\end{equation}
The corresponding field equation is 
\begin{equation}
\Box \phi +[-J_\mu (\phi )J^\mu (\phi )-l_0^{-1}N_\mu J^\mu (\phi
)]\phi =0.
\label{eq6}\end{equation}

Our basic observation is that in this model the source of the divergence is
invariant under the duality transformation
\begin{equation}
\label{eq7}\phi \rightarrow \phi ^{-1},\hspace{5mm}N_\mu \rightarrow -N_\mu, 
\end{equation}
which connects the interchange of $\phi $ and its reciprocal value $\phi ^{-1}$
to the reversal of the direction of the internal vector. This duality
transformation, which can alternatively be viewed as reflecting 
the invariance property of the dynamical mass term in (\ref{eq6}), can be
generalized to a dynamical symmetry of the entire field equation (\ref{eq6}). To this end, we first work out how the left hand side of (\ref{eq6})
transforms under (\ref{eq7}). We find 
\begin{eqnarray}
\Box\phi^{-1}&+&[-J_\mu(\phi^{-1})J^\mu(\phi^{-1})-l_0^{-1}(-N_\mu)J^\mu
(\phi^{-1})]\phi^{-1}\nonumber\\
&=&-\phi^{-2}\{\Box\phi+[-J_\mu(\phi)J^\mu(\phi)+l_0^{-1}N_\mu
J^\mu(\phi)]\phi\},     \label{eq8}
\end{eqnarray}
from which one infers
that if $(\phi ,N_\mu )$ is a solution of (\ref{eq6}),
then $(\phi^{-1},-N_\mu )$
can not be a solution.
Under the assumption that $(\phi ,N_\mu )$ is a solution of (\ref{eq6})
we get 
for $(\phi^{-1},-N_\mu)$ the equation
\begin{equation}
\Box\phi^{-1}+[-J_\mu(\phi^{-1})J^\mu(\phi^{-1})+l_0^{-1}(-N_\mu)J^\mu
(\phi^{-1})]\phi^{-1}=0.\label{8b}
\end{equation}
The difficulty here is that the term
$l_0^{-1}N_\mu J^\mu (\phi )$ on the left hand side, if compared with
the corresponding term on the left hand side of (\ref{eq6}),
appears with a wrong sign.

In order to obtain a dynamical symmetry for the dual
configurations $(\phi ,N_\mu )$ and $(\phi^{-1},-N_\mu )$,
we generalize the duality transformation in such a way as to involve a
transition from a Lorentzian signature $(-,+++)$ to a Euclidean one. Due to
the assumed time-like nature of the internal vector on a Lorentzian
domain (a domain with Lorentzian metric), the signature transition would
then compensate the wrong sign of
$l_0^{-1}N_\mu J^\mu (\phi )$ on the 
left hand side of (\ref{8b})\footnote{We note that 
the identity (\ref{eq4}) holds no matter what signature is used
in its derivation.}.\\ To formulate this desired form of duality
transformation we first implement the transition from the Lorentzian
signature to the Euclidean one in an intrinsic way by the
metric-transformation 
\begin{equation}
\eta _{\mu \nu }\rightarrow \eta _{\mu \nu }+2N_\mu N_\nu =:\bar \eta _{\mu
\nu}.
\end{equation}
Clearly, the transformed metric $\bar \eta _{\mu \nu }$ has a Euclidean
signature. The duality transformation we wish to consider must then take the
form 
\begin{equation}
\label{D}(\phi )\rightarrow (\phi )^D,
\end{equation}
where we have used the abbreviation $(\phi )=(\phi ,N_\mu ,\eta _{\mu \nu })$
and $(\phi )^D=(\phi ^{-1},-N_\mu ,\bar \eta _{\mu \nu })$. We emphasize
that the field equation (\ref{eq6}) makes no distinction between $(\phi )$
and its dual $(\phi )^D$. In particular, we can make no distinction between
a Lorentzian and an Euclidean signature in terms of the field equation. In
fact, both signatures are related by the dynamical symmetry of the field
equation whose transformations reverse the value of the scalar field $\phi $
and the direction of the internal vector at each point.
 This aspect reflects one
of the characteristic features of the divergence theory (\ref{eq5}).

\section{The time-asymmetric condition of duality-breaking}
At the dynamical level there is no distinction between $(\phi )$ and its
dual $(\phi )^D$. One may take $(\phi )$ and $(\phi )^D$ as equivalent
configurations in vacuum. However, a distinction between $(\phi )$ and $%
(\phi )^D$ can actually be made if one takes into account the boundary
conditions to be imposed on the physically realizable configuration of
vacuum. One possibility is to subject the choice of the boundary conditions
to external limitations imposed on the large scale characteristics
 of the scalar
field $\phi $. Such limitations would affect the background average value of $%
\phi$, with the  obvious consequence that the dynamically allowed
transition between dual configurations $(\phi )$ and $(\phi )^D$ can be
excluded.

For many reasons, it may be of interest to realize a
duality-breaking directly by means of boundary conditions. But our approach
here is different. It is based on the realization that a duality-breaking,
irrespective of its particular realization, must be connected with the
emergence of a preferred arrow of time in vacuum, provided we focus
ourselves to a Lorentzian domain. In fact, on such a domain any preferred
configuration of the scalar field $\phi $ selected by the condition of the
duality-breaking has to select a preferred time-direction for the internal
vector $N_\mu $. This is clearly reflected in the form of
duality-transformation (\ref{D}) which combines the transformation-law 
of $\phi $ and $N_\mu $in an essential way. 
We shall take this mutual interrelation between
duality-breaking and time-asymmetry as the basic input for expressing the
condition of the duality-breaking on a Lorentzian domain. In explicit terms,
our approach is to combine the divergence theory (\ref{eq5}) with a
time-asymmetric law on a Lorentzian domain, formulated as the statement 
\begin{eqnarray}
\partial_\mu J^\mu(\phi)=l_0^{-1}N_\mu J^\mu(\phi) > 0. 
\label{db}
\end{eqnarray}
Since by the duality transformations (\ref{D}) the source of the divergence
in (\ref{eq5}) will reverse the sign, the condition (\ref{db}) can not
simultaneously be satisfied for solutions related by duality
transformations. Therefore, the law (\ref{db}) does imply an assertion about
both the violation of duality and the asymmetry in the direction of the
internal vector $N_\mu $ in vacuum. This asymmetry shows a general tendency
to establish a preferred time-direction for $N_\mu $.

\section{Vacuum-condensation}

The duality breaking law (\ref{db}) ensures that on a Lorentzian domain the
source of the divergence has a tachionic contribution to the dynamical mass
term in (\ref{eq4}). This provides the indication that the law (\ref{db})
may act to produce the condensation of vacuum. In this section, we shall
establish this possibility in explicit terms. 

If one takes the divergence
law (\ref{eq5}) together with (\ref{eq1}) one finds 
\begin{eqnarray}
\partial_\mu J^\mu(\phi)=l_0^{-1}N_\mu\partial^\mu\ln\phi,
\label{eq10}
\end{eqnarray}
which has a particularly simple solution, namely 
\begin{equation}
\label{eq11}J^\mu (\phi )=l_0^{-1}N^\mu \ln \phi. 
\end{equation}
Restricting ourselves to this solution, we can write the divergence law
(\ref{eq5}) as
\begin{eqnarray}
\partial_\mu J^\mu(\phi)=-l_0^{-2}\ln\phi,  \label{eq12}
\end{eqnarray}
which, if combined with the condition (\ref{db}) could then interpret
those configurations of the scalar field $\phi $ as physical for which the
value of $\phi $ remains smaller than one. In general, a condition of the
type 
\begin{equation}
\phi (x)<1,
\label{F}\end{equation}
as suggested by the condition of duality breaking, characterizes the physical
realizability of the field $\phi$
on a Lorentzian domain. Using (\ref{eq12}) we may
write the field equation (\ref{eq6}) in the form 
\begin{equation}
\Box\phi+l_{0}^{-2}\left[(\ln\phi)^2+\ln\phi\right]\phi=0.
\label{eq13}\end{equation}
The last equation can be derived from an effective Lagrangian 
\begin{eqnarray}
{\cal{L}}~\sim~ \frac{1}{2}\left(\partial_\mu\phi\partial^\mu
\phi-V(\phi)\right),  \label{la}
\end{eqnarray}
where 
\begin{eqnarray}
V(\phi)=\phi-\phi\ln \phi+\phi(\ln\phi)^2. \label{pot}
\end{eqnarray}
This potential admits a condensation of vacuum associated with the 
minimum of potential (\ref{pot}) which is located at 
the constant value $\phi=1$. 

Remarkably,
there exists a deep reason for ragarding the constant 
value $\phi=1$ as a disreable
characteristic for the vacuum-condensation in the present case.
In fact, it can generally be expected that
a universal form of time-asymmetry in vacuum is not detectable
by a constant ground-state configuration. Rather,
it can only be detected in the presence of a
systematic form of fluctuations around a given ground-state configuration. 
In this sense, the value
$\phi=1$ appears to be very desireable because the corresponding 
ground-state configuration excludes the universal form of time-asymmetry 
reflected in (\ref{db}), because 
the source of the divergence in (\ref{db}) vanishes.
	 
In the presence of small deviations 
of $\phi$ from the ground state $\phi=1$, it is
possible to derive an approximative form for the potential.
For this purpose we may first use a linear approximation 
 for $\ln\phi$ in (\ref{eq13}), namely
\begin{eqnarray}
\ln\phi\approx -1+\phi,
\label{eq14}
\end{eqnarray}
which corresponds to the first terms in a Taylor expansion of $\ln \phi $
around the minimum $\phi=1$. 
Using the last relation, we may
write equation (\ref{eq13}) as 
\begin{eqnarray}
\Box\phi+l_0^{-2} [-\phi+\phi^{2}]\phi=0.
\label{eq15}
\end{eqnarray}
An effective form of this equation can be obtained if in the dynamical
mass-term we approximate the linear term in $\phi$ by the coresponding
ground-state value to
obtain 
\begin{eqnarray}
\Box\phi+l_0^{-2}[-1+\phi^{2}]\phi=0.
\end{eqnarray}
The last equation may now be derived from  Lagrangian (\ref{la}) with 
\begin{eqnarray}
V(\phi)=-l_0^{-2}\phi^2+\frac{1}{2}l_0^{-2}\phi^4, 
\nonumber
\end{eqnarray}
which has the form of the Higgs potential. We emphasize that this form
holds only for small deviations from the ground state-value of the potential
(\ref{pot}).

\section{Concluding remarks}

The paper has studied the consequence of the violation of Lorentz-invariance
in vacuum in a simple model. One of the characteristic features of this
model is that the requirement of a fixed background signature does not
correspond to a property demanded by the dynamical laws. Indeed, we have
derived the signature changing duality transformations as a natural symmetry
transformations of the model. It is only if the dynamical laws are combined
with a duality-breaking law in vacuum that a fixed background signature can
be established within the model. If the latter signature is taken to be the
Lorentzian one, the duality-breaking law then implies an assertion about the
emergence of a preferred arrow of time, namely that determined by the direction
of the internal vector. In this respect the duality-breaking is linked with
an inherent time-asymmetry in vacuum. It was shown that this
time-asymmetry can act to produce a corresponding condensation of vacuum.\\
A final remark may be of interest. 
It is possible that the condition (\ref{F}) 
which was found to characterize the physical realizability of the field 
$\phi $, as demanded by the duality-breaking law, may not be satisfied for a
solution on the entire Einstein-Minkowski space. If this would happen then
$\phi (x)=1$ would characterize a signature-changing hypersurface in that
space. On such a hypersurface the scalar field $\phi $ has a self-dual
configuration associated with its ground-state configuration in vacuum  
and the source of the divergence in (\ref{eq11}) vanishes (no fluctuations
on the hypersurface).
These features would constraint the form of physical      
junction-conditions for relating the
solutions across the hypersurface. In disjoint regions next to the
hypersurface the solutions would then be related by a duality transformation.
This remark may be of interest to examine nature of the signature-changing
hypersurface and the corresponding junction conditions not only in the
present context, but also in the gravitational context where various models
describing a signature transition have been found, see e.g. \cite{dereli}.\vspace{10mm}\\
{\bf Acknowledgment}\vspace{3mm}\\
The authours wish to thank R W Tucker for a useful discussion regarding a 
previous version of this paper.


\begin{thebibliography}{99}
\bibitem{blok} Blokhintsev d I, 1964 Physics Letters 12, 272.
\bibitem{nielsen} Nielsen H B and Picek I, 1983 Nuclear Physics B 211, 269.
\bibitem{phil} Phillips P R, 1965 Physical Review 139(2B), 491.
\bibitem{salehi}  Salehi H, International Journal of Theoretical Physics 1997, 39, 9, 2035.
\bibitem{dereli} Dereli T and Tucker R W, 1993 Class. Quantum Grav. 10, 365.
\end{thebibliography}
\end{document}